\documentclass[letterpaper, 10 pt, conference]{ieeeconf}  

\IEEEoverridecommandlockouts                              
\usepackage{graphics} 
\usepackage{epsfig} 
\usepackage{subcaption}
\usepackage[noadjust]{cite}
\usepackage{amsmath,amssymb,amsfonts,mathrsfs} 
\usepackage{algorithm,algpseudocode}
\usepackage{balance}
\usepackage{tcolorbox}
\usepackage{bbm} 

\usepackage{todonotes}
\usepackage{soul}
\definecolor{smoothgreen}{rgb}{0.7,1,0.7}
\sethlcolor{smoothgreen}
\newcommand{\todopara}[1]{\vspace{0px} %
\todo[inline, color=black!10]{\textbf{[Paragraph:]} {#1}} %
}

\newcommand{\todonote}[1]{\vspace{0px} %
	\todo[inline, color=green!30]{\textbf{[Note:]} {#1}} %
}

\newcommand{\todoQ}[1]{\vspace{0px} %
	\todo[inline, color=orange!50]{\textbf{[Note:]} {#1}} %
}

\newcommand{\todohere}[1]{\hl{(\textbf{TODO:} #1)}}
\newcommand{\hidetodos}{
	\renewcommand{\todopara}[1]{}
	\renewcommand{\todonote}[1]{}
	\renewcommand{\todoQ}[1]{}
	\renewcommand{\todohere}[1]{}
}

\newcommand{\clnote}[1]{\ifthenelse{\boolean{include-notes}}%
	{\textcolor{orange}{\textbf{CL: #1}}}{}}

\title{\LARGE \bf
Rumor-robust Decentralized Gaussian Process Learning, Fusion, and Planning for Modeling Multiple Moving Targets}

\author{Chang Liu$^\dag$ \textit{Member, IEEE}, Zhihao Liao$^\ddag$, and Silvia Ferrari$^\S$ \textit{Senior Member, IEEE}
	\thanks{$^\dag$Chang Liu is a Software Engineer at Nvidia Corp. He was a Postdoctoral Associate at Sibley School of Mechanical and Aerospace Engineering at Cornell University. Email: {\tt\small changliu1289@gmail.com}.}
	\thanks{$^\ddag$Zhihao Liao is a Software Engineer at TuSimple. He was a Master student at Sibley School of Mechanical and Aerospace Engineering at Cornell University. Email: {\tt\small zl673@cornell.edu}.}
	\thanks{$^\S$Silvia Ferrari is the John Brancaccio Professor of Mechanical and Aerospace Engineering at Cornell University, Ithaca, NY USA. Email: {\tt\small sf375@cornell.edu}.}
}

\begin{document}

\maketitle
\hidetodos
\thispagestyle{empty}
\pagestyle{empty}

\begin{abstract}
	This paper\footnote{A short version of this paper is published in 59th IEEE Conference on Decision and Control, 2020.} presents a decentralized Gaussian Process (GP) learning, fusion, and planning (RESIN) formalism for mobile sensor networks to actively learn target motion models. RESIN is characterized by both computational and communication efficiency, and the robustness to rumor propagation in sensor networks. By using the weighted exponential product rule and the Chernoff information, a rumor-robust decentralized GP fusion approach is developed to generate a globally consistent target trajectory prediction from local GP models. A decentralized information-driven path planning approach is then proposed for mobile sensors to generate informative sensing paths. A novel, constant-sized information sharing strategy is developed for path coordination between sensors, and an analytical objective function is derived that significantly reduces the computational complexity of the path planning. The effectiveness of RESIN is demonstrated in numerical simulations.
\end{abstract}
\section{INTRODUCTION}
The problem of learning the behavior and dynamics of moving targets via mobile sensor networks has received significant attention in recent years because of important applications such as environmental monitoring \cite{singhModelingEnvironmentalSurveillance10,liuDBFLIFO17}, security and surveillance \cite{liuDBF18,weiAutoCameraControlForLearningDPGP18,foderaroDOCTargetTracking18}, and the internet of things \cite{hongSnailIoT10}. 
In these applications, the region of interest (ROI) greatly exceeds the size of the sensor's field of view (FOV) and, therefore, managing mobile sensor positions is crucial to collectively learn the motion models of targets in the ROI. 

Bayesian nonparametric (BNP) models, such as Gaussian Processes (GPs) and Dirichlet Process Gaussian Processes (DPGPs), have been shown very effective at modeling moving targets because of their flexibility, expressiveness, and data-driven nature \cite{rasmussenGaussianProcessforMachineLearning06}. Unlike traditional, model-based approaches, GPs require little prior information about target behavior, and are applicable when the number of targets of interest change over time, for example, as new targets enter and old targets leave the ROI \cite{weiGeometricTransversalTracking15,weiAutoCameraControlForLearningDPGP18,liuLearnRecurBNP19}.
As a result, GP models provide a more flexible and systematic approach for modeling moving targets when compared to semi-Markov jump systems \cite{luInformationPotentialTracking11}, linear stochastic models \cite{martinezOptimalSensorPlacement06}, and physics-based models \cite{leeParallelIMMMotionPrediction17,liKFTrackingUltrasonicSensorArray18}. Early works on BNP sensor network control relied on centralized learning, data fusion and planning \cite{osborneTowardsRealtimeInfoSensorNetworkMultioutputGP08,xuMobileSenosrGPTruncatedObservation11}. 
In many applications, however, contested communication and GPS-denied environments prevent centralized methods from performing robustly and reliably. 
This is because the central station or fusion agent may be unable to gather information and/or convey plans to all sensors consistently over time.
This paper presents a decentralized BNP learning, data fusion, and planning formalism characterized by easiness for parallelization, scalability, and robustness to single-point failure compared to the centralized counterparts. 

Several decentralized GP learning and fusion approaches have been previously developed in the literature to achieve computations that are distributed among independent local agents operating on subsets of the data. 
Two representative classes of methods include the mixture of experts (MoEs) \cite{yukselTwentyYearsMixtureExperts12} and the product of experts (PoEs) \cite{caoGeneralizedPoEGPfusion14,deisenrothDistributedGP15}.
In MoEs, each agent locally learns a GP model for a different partition of the state space and the global prediction is made by collecting all of agents' local predictions. 
A gating network is used to assign weights to each local prediction based on the corresponding agent's domain of training data. 
In contrast, in PoEs, agents share the same state space and each agent locally learns an independent GP model using a subset of training data. 
Then, the global prediction is made by the Bayes rule and the independence assumption of local predictions. 
PoE methods allow for efficient training and prediction and thus have attracted great interest.
However, current PoE approaches cannot be directly applied to data fusion in sensor networks since PoE-based approaches cannot handle rumor propagation. 
This means common information between local agents, such as the simultaneous measurements of the same target, may be redundantly used and leads to incorrect fusion results \cite{campbellDistributedDataFusion16}. 
This paper presents a ``rumor-robust'' PoE-based approach for fusing GP models such that rumor propagation is prevented. 

A decentralized information-driven path planning (IPP) approach is also presented for controlling and coordinating sensor trajectories so as to obtain the most informative target measurements subject to communication constraints. 
Previous methods for decentralized IPP include 
a decentralized, gradient-based control approach that assumes all-to-all agent communications \cite{hoffmannMobileSensorNetwork10}. 
By this approach, the sensor measurements and the gradient of their objective functions are communicated constantly with the network, and the trajectories are proven to converge to a Nash equilibrium. Since the convergence of gradient-based optimization requires multiple iterations, this method incurs large communication burden. 
A decentralized planning and adaptive grouping method is presented in \cite{ouyangMultirobotGP14}. The local planning of sensors is conducted in a centralized manner where a local fusion center computes the optimal trajectories for all sensors in the group. While the planning is fully decentralized, there is no performance guarantee of the planning result. 
More recently, \cite{liMultirobotOnlineSensingCommMap19} proposes a multi-robot online sensing strategy for the construction of communication maps using GP. The work uses a leader-follower paradigm, where each pair of leader-follower sensors is manually defined and the leader generates plans to coordinate with its follower. However, no coordination is ensured between different pairs. 

To overcome these challenges, this paper develops a rumor-robust decentralized GP learning, fusion, and planning (RESIN\footnote{RESIN is the acronym of ``Rumor-robust decentralized gp lEarning, fuSIon, and planNing''.}) approach for mobile sensor networks to actively learn target motion models. 
To deal with time-varying target motion models, a spatio-temporal kernel is used in GP modeling. A rumor-robust decentralized GP learning and fusion algorithm is then proposed and applied to combine individual sensors' local prediction of target trajectories into a globally consistent one. The GP learning and fusion approach is computationally efficient and can avoid rumor propagation in the sensor network. 
We subsequently present a 
sequential optimal control approach that is efficient both in computation and communication for decentralized sensor path planning. 
In particular, an analytical objective function is derived via the use of decentralized GP fusion, which reduces the original mixed integer nonlinear programming problem into a low-dimensional nonlinear programming problem. Besides, a new information sharing strategy is proposed for coordination between sensors, which only requires a constant-sized communication overhead. In contrast, prior works \cite{krauseNearOptObserveSelectSubmodFunc07,atanasovDecentralizedActiveInfo15,damesDetectingLocalizingTrackingFISST17} require a communication overhead that is linear in the number of sensors. 

The rest of the paper is organized as follows. The problem formulation is presented in Section \ref{sec:prob_form}. The decentralized GP learning and fusion approach is described in Section \ref{sec:dec_gp}. Section \ref{sec:dec_ctrl} subsequently presents the decentralized path planning algorithm. Numerical simulation results are presented in Section \ref{sec:sim}. Conclusions are drawn in Section \ref{sec:conclusion}.

\section{Problem Formulation} \label{sec:prob_form}
\begin{figure}
    \includegraphics[width=0.45\textwidth]{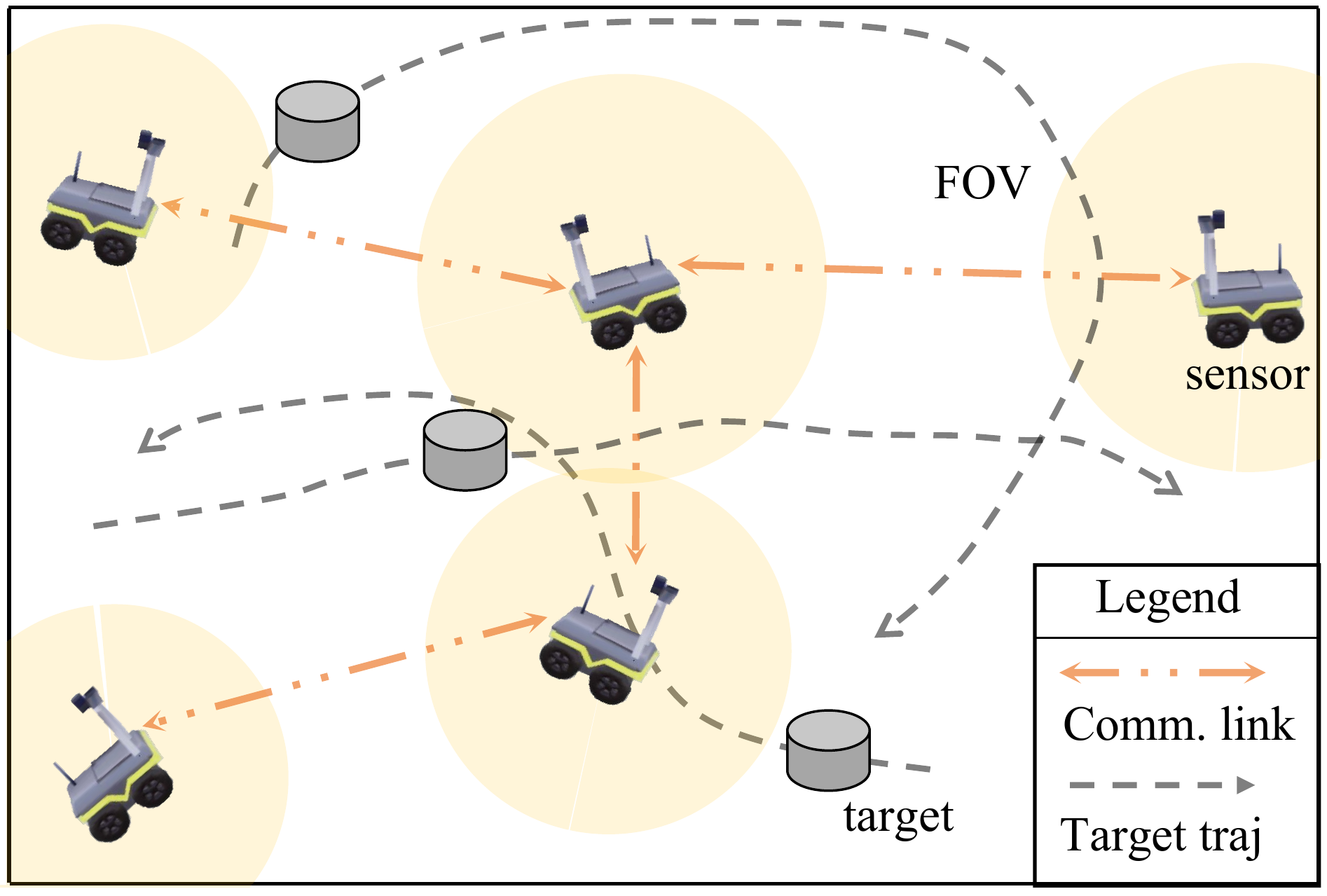}
    \centering
    \caption{Sensors share information between neighbors and coordinate their trajectories to actively learn the GP motion models of multiple moving targets.}
\label{fig:setup}
\end{figure}

Consider a network of $N$ mobile sensors deployed to learn the motion models of $M$ targets moving across a connected, compact, and non-empty workspace $\mathcal{W}\subset\mathbb{R}^{2}$ (Figure \ref{fig:setup}). 
Each sensor is equipped with a fixed stereo-camera and a wireless communication device. 
The number of targets is unknown a priori and can change over time due, for example, to targets entering or exiting the workspace. 
The motion model of each target, indexed by $i$, is represented by a time-varying, continuous, and differentiable function $\mathbf{f}_{i}:\mathbb{R}^{2}\times\mathbb{R}\rightarrow\mathbb{R}^{2}$ defined over $\mathcal{W}$, which maps the target position to its velocity, i.e.,
\begin{equation}
    \label{eqn:tar_kin_model}
    \dot{\mathbf{x}}_{i}(t)=\mathbf{f}_{i}\left[\mathbf{x}_{i}(t),t\right]\triangleq\mathbf{v}_{i}(t),
\end{equation}
where $\mathbf{x}_{i}(t)\in\mathcal{W}$ and $\mathbf{v}_{i}(t)\in\mathbb{R}^{2}$ represent the position and velocity of the target's center of mass, respectively.

The sensor's state, defined as $\mathbf{s}=[s_{x}\quad s_{y}\quad s_{\theta}\quad s_{v}]^{T}\in\mathbb{R}^{4}$, includes the sensor position $[s_{x}\quad s_{y}]^T\in\mathcal{W}$, orientation $s_{\theta}\in[0,2\pi)$, and velocity $s_{v}\geq0$. 
The sensor control input, defined as $\mathbf{u}=[a\quad\omega]^{T}\in\mathbb{R}^{2}$, includes the linear acceleration, $a\in\mathbb{R}$, and angular velocity, $\omega\in\mathbb{R}$.
Let $\Delta T>0$ represent the discretization interval so that the $k$th step corresponds to $t=k\Delta T$. The $j$th sensor's kinematic model can be represented by the following difference equation, 
\begin{equation}
\mathbf{s}_{j}(k+1)=\mathbf{s}_{j}(k)+\begin{bmatrix}s_{v}(k)\cos(s_{\theta}(k))\\
s_{v}(k)\sin(s_{\theta}(k))\\
\mathbf{u}_{j}(k)
\end{bmatrix}\Delta T\label{eqn:sensor_kin_model}
\end{equation}
For simplicity, in the rest of the paper, we refer to the kinematic
model (\ref{eqn:sensor_kin_model}) as $\mathbf{g}:\mathbb{R}^{4}\times\mathbb{R}^{2}\rightarrow\mathbb{R}^{4}$
such that 
\begin{equation}
\mathbf{s}_{j}(k+1)=\mathbf{g}(\mathbf{s}_{j}(k),\mathbf{u}_{j}(k)).
\end{equation}

Each sensor measures the target position and velocity by computing the sparse scene flow of the target \cite{lenzSparseSceneFlowSeg11}. 
Camera frames containing targets are obtained when the targets are inside the sensor's FOV, defined as $\mathcal{F}\left(\mathbf{s}_{j}(k)\right)=\left\{ \mathbf{w}\in\mathcal{W}~|~\left\Vert [s_x(k)\quad s_y(k)]^T-\mathbf{w}\right\Vert _{2}\leq r_{j}\right\} $, where $r_{j}>0$ denotes the $j$th sensor's sensing range. 
The camera obeys the following measurement model with additive Gaussian noise, 
\begin{equation}
\mathbf{z}_{ij}(k)=\begin{cases}
\mathbf{v}_{i}(k)+\boldsymbol{\varepsilon} & \text{if }\mathbf{x}_{i}(k)\in\mathcal{F}\left(\mathbf{s}_{j}(k)\right)\\
\emptyset & \text{if }\mathbf{x}_{i}(k)\notin\mathcal{F}\left(\mathbf{s}_{j}(k)\right)
\end{cases},\label{eqn:msmt_model}
\end{equation}
where $\mathbf{z}_{ij}(k)\in\mathbb{R}^{2}$ is the $j$th sensor's
measured velocity of $i$th target, and $\boldsymbol{\varepsilon}\in\mathbb{R}^{2}$
is a zero-mean Gaussian white noise and it follows the distribution $\mathcal{N}(\mathbf{0},\boldsymbol{\Sigma}_\varepsilon)$, where $\boldsymbol{\Sigma}_\varepsilon=\epsilon_{0}^{2}\mathbf{I}$.
Here we assume perfect data association between targets and sensor measurements, as the data association problem is out of the scope of this paper.

The sensors form a communication network where each sensor can constantly communicate with its neighboring sensors. 
Due to the limited communication range in practical applications, sensors form local groups where sensors within the same group can communicate with each other via single or multiple hops, i.e., the communication network forms a connected graph \cite{mesbahiGraphTheoryMethodMultiagentNetwork10}. 
RESIN is run within each group. As it will become clear in later sections, RESIN can well handle the dynamic change of grouping over time. Therefore, without loss of generality, in the rest of the paper, it is assumed that all $N$ sensors form a single group 
and a tree-structured communication path always exists at each time step.

\subsection{Decentralized Learning and Fusion}
In order to learn the target motion models, sensors must accurately predict target positions and actively decide their sensing trajectories to reduce the uncertainty in the target states estimates. 
Each sensor first locally learns a GP model based on its own sensor measurements to predict the targets' future trajectories and then fuses the local prediction into a global one via communicating with neighboring sensors. 
A typical issue in decentralized fusion is the rumor propagation, where the common information in sensors' local data are double counted. 
To see this, let $P_{j}\left(\mathbf{x}(k)~|~\mathbf{Z}_{j}(k)\right)$ and $P_{l}\left(\mathbf{x}(k)~|~\mathbf{Z}_{l}(k)\right)$ represent the probability density function (pdf) of local estimates of $\mathbf{x}(k)$ from sensors $j$ and $l$, respectively, and let $\mathbf{Z}_{j}$ and $\mathbf{Z}_{l}$ represent the corresponding collection of measurements by sensors $j$ and $l$.
Then the fused pdf is \cite{ahmedFastChernoofFusionGM12}
\begin{multline}
P\left(\mathbf{x}(k)~|~\mathbf{Z}_{j}(k)\cup\mathbf{Z}_{l}(k)\right)\\\propto\frac{P\left(\mathbf{x}(k)~|~\mathbf{Z}_{j}(k)\right)P\left(\mathbf{x}(k)~|~\mathbf{Z}_{l}(k)\right)}{P\left(\mathbf{x}(k)~|~\mathbf{Z}_{j}(k)\cap\mathbf{Z}_{l}(k)\right)},
\end{multline}
where the denominator is the conditional probability distribution based on common information between sensors $j$ and $l$. Tracking and removing the common information is needed to avoid rumor propagation, but is usually computationally heavy. This paper proposes a decentralized GP fusion approach to combine local predictions to generate globally consistent prediction of target trajectories while avoiding double counting.

\subsection{Information-driven Path Planning (IPP) Algorithm}
The IPP can be formulated as an optimal control problem. 
Let $\boldsymbol{u}_j\left(k:k_{f}\right)=\left[\boldsymbol{u}_j^{T}\left(k\right)\quad\dots\quad\boldsymbol{u}_j^{T}\left(k_{f}\right)\right]^{T}$ represent the planned control inputs of $j$th sensor over the planning interval $[k,k_{f}]$ and $\boldsymbol{U}\left(k:k_{f}\right)=\left[\boldsymbol{u}_{1}\left(k:k_{f}\right)\quad\dots\quad\boldsymbol{u}_{N}\left(k:k_{f}\right)\right]$ denote the control inputs of all $N$ sensors. The optimal control of all sensors, $\boldsymbol{U}^{*}\left(k:k_{f}\right)$, is computed by maximizing the objective function $J\left(\boldsymbol{U}\left(k:k_{f}\right)\right)$ under system constraints, formulated as the following optimization problem,
\begin{equation}
\begin{aligned}
\label{eq:cent_ctrl}
\boldsymbol{U}^{*}\left(k:k_{f}\right) & =\arg\max_{\boldsymbol{U}\left(k:k_{f}\right)}J\left(\boldsymbol{U}\left(k:k_{f}\right)\right)\\
 s.t. ~~& \mathbf{s}_j(\tau+1)=\mathbf{g}\left(\mathbf{s}_j(\tau),\mathbf{u}_{i}(\tau)\right)\\
 & \mathbf{s}_j(\tau)\in\mathcal{S},\;\mathbf{u}_j(\tau)\in\mathcal{U}\\
 & \tau=k,\dots,k_{f}-1,\,j=1,\dots,N,
\end{aligned}
\end{equation}
where $\mathcal{S}$ and $\mathcal{U}$ represent the feasible set of sensor state and control input, respectively. 

In this work, the objective function uses the mutual information (MI),
which has been shown very effective for information-driven path planning \cite{hoffmannMobileSensorNetwork10,weiInformationValueDPGP16,weiAutoCameraControlForLearningDPGP18}. In particular, define $\mathbf{X}\left(k:k_{f}\right)=\left[\mathbf{x}_{1}\left(k:k_{f}\right)\quad\dots\quad\mathbf{x}_{m}\left(k:k_{f}\right)\right]$ as the predicted target positions in the planning interval, which are obtained using decentralized GP fusion. Also define $\mathbf{Z}\left(k:k_{f}\right)$ as the predicted measurements of these targets from all sensors, then $J\left(\boldsymbol{U}\left(k:k_{f}\right)\right)$ is defined as the conditional MI between $\mathbf{X}\left(k:k_{f}\right)$ and $\mathbf{Z}\left(k:k_{f}\right)$, given existing sensor measurements, i.e.,
\begin{equation}
\label{eq:ctrl_obj}
J\left(\boldsymbol{U}\left(k:k_{f}\right)\right)=I\left(\mathbf{X}\left(k:k_{f}\right);\mathbf{Z}\left(k:k_{f}\right)~|~\mathbf{Z}\left(1:k-1\right)\right).
\end{equation}
Note that, though $\boldsymbol{U}\left(k:k_{f}\right)$ does not explicitly appear in (\ref{eq:ctrl_obj}), the sensor control directly decides the planned sensor positions $\mathbf{S}\left(k:k_{f}\right)$ and, consequently, the expected sensor measurements $\mathbf{Z}\left(k:k_{f}\right)$, thus determining the objective function.

Solving the centralized IPP (\ref{eq:cent_ctrl}) is in general computationally expensive due to the exponentially growing search space with respect to the sensor number and the planning horizon. This paper propose a decentralized IPP algorithm to distribute the computation among sensors to efficiently solve the problem.

\section{Decentralized GP Learning and Fusion for Target Prediction}\label{sec:dec_gp}
The GP modeling approach adopted in this paper is a Bayesian non-parametric approach that has gained increased popularity in recent years for learning system dynamics from data \cite{liuGPLearnCtrlTutorial18,weiAutoCameraControlForLearningDPGP18,liuLearnRecurBNP19}. 
A GP defines a distribution over functions and the function values at any finite set of input points form a joint Gaussian distribution.
This works uses GP to model the velocity field of each target, with the target position and time being the input and the corresponding target velocity being the output of GP. 
In particular, let $\mathbf{X}_{i}(k)=\left[\mathbf{x}_{i}(1)\quad\dots\quad\mathbf{x}_{i}(k)\right]$ and $\mathbf{Z}_{ij}(k)=\left[\mathbf{z}_{ij}(1)\quad\dots\quad\mathbf{z}_{ij}(k)\right]$ represent the measured positions and velocities of $i$th target by the $j$th sensor at time steps $1$ to $k$.
Then $j$th sensor's local GP model can predict the velocity given a query position $\boldsymbol{\xi}\in\mathcal{W}$ and time $\tau>0$, with the predicted value $\mathbf{z}_{ij,\boldsymbol{\xi}}(\tau)$ obeying the following Gaussian distribution \cite{rasmussenGaussianProcessforMachineLearning06}, 
\begin{equation}
\begin{aligned}\mathbf{z}_{ij,\boldsymbol{\xi}}(\tau) & \sim\mathcal{N}\left(\boldsymbol{\mu}_{ij}(\boldsymbol{\xi}),\Sigma_{ij}(\boldsymbol{\xi})\right)\\
\boldsymbol{\mu}_{ij}(\boldsymbol{\xi}) & =\mathbf{K}\left(\boldsymbol{\xi},\mathbf{X}\right)\left(\mathbf{K}\left(\mathbf{X},\mathbf{X}\right)+\epsilon_{0}^{2}\mathbf{I}\right)^{-1}\mathbf{Z}\\
\Sigma_{ij}(\boldsymbol{\xi}) & =\mathbf{K}\left(\boldsymbol{\xi},\boldsymbol{\xi}\right)\\
&~~~~~~-\mathbf{K}\left(\boldsymbol{\xi},\mathbf{X}\right)\left(\mathbf{K}\left(\mathbf{X},\mathbf{X}\right)+\epsilon_{0}^{2}\mathbf{I}\right)^{-1}\mathbf{K}\left(\mathbf{X},\boldsymbol{\xi}\right)
\end{aligned}
\label{eqn:gp_formula}
\end{equation}
where $\boldsymbol{\mu}_{ij}(\boldsymbol{\xi})$ and $\Sigma_{ij}(\boldsymbol{\xi})$ represent the mean and covariance matrix of the Gaussian distribution.
The kernel matrix $\mathbf{K}(\cdot,\cdot)$ is the key component of GP and it encodes the similarity between input data points. 
This paper uses the following spatial-temporal Radial Basis functions as the spatio-temporal kernel to account for the time-varying nature of the motion model,
\begin{equation*}
K\left(\mathbf{x}_{i}(t_{i}),\mathbf{x}_{j}(t_{j})\right)=\sigma_{s}^{2}e^{-\frac{\left\Vert \mathbf{x}_{i}-\mathbf{x}_{j}\right\Vert _{2}^{2}}{2l_{x}^{2}}}e^{-\frac{\left(t_{i}-t_{j}\right)^{2}}{2l_{\tau}^{2}}},
\end{equation*}
where $l_{x}$ and $l_{\tau}$ represent the spatial and temporal length scale, respectively, and $\sigma_{s}$ is the hyperparameter for signal variance.

\subsection{Local GP Learning and Prediction of Target Trajectory}\label{subsec:local_gp_pred}
Given the measurements of sensor $j$, the hyper-parameters of the local GP can be learned by maximizing the logarithm of the marginal likelihood function of the training data \cite{rasmussenGaussianProcessforMachineLearning06}. 
The resultant GP model is then used to predict the target positions in the planning interval $\left[k,k_{f}\right].$ Using the Bayes rule, the pdf of predicted target positions can be represented as follows,
\begin{equation}
	\begin{aligned}
		&P_j\left(\mathbf{X}_{i}(k+1:k_{f})~|~\mathbf{X}_{i}(k)\right)=\prod_{\tau=k}^{k_{f}-1}P\left(\mathbf{x}_{i}(\tau+1)~|~\mathbf{x}_{i}(\tau)\right)\label{eqn:local_gp_pred}\\
		&~~~=\prod_{\tau=k}^{k_{f}-1}\int_{\mathbb{R}^{2}}\left[P\left(\mathbf{x}_{i}(\tau+1)~|~\mathbf{v}_{i}(\tau),\mathbf{x}_{i}(\tau)\right)\right.\\
		&~~~~~~~~~~\left.P_j\left(\mathbf{v}_{i}(\tau)~|~\mathbf{x}_{i}(\tau)\right)\right]d\mathbf{v}_{i}(\tau),
	\end{aligned}
\end{equation}
where $P_j\left(\mathbf{v}_{i}(\tau)~|~\mathbf{x}_{i}(\tau)\right)$ corresponds to $j$th sensor's local GP model and $P\left(\mathbf{x}_{i}(\tau+1)~|~\mathbf{v}_{i}(\tau),\mathbf{x}_{i}(\tau)\right)$ can be obtained from the target motion model (\ref{eqn:tar_kin_model}).
The factorizatoin in (\ref{eqn:local_gp_pred}) is due to the Markov property of the target motion model. 

In general, there is no analytical form for $P_j\left(\mathbf{X}_{i}(k+1:k_{f})~|~\mathbf{X}_{i}(k)\right)$ when $k_{f}-k\ge2$. 
To make the prediction tractable, we define a \textit{nominal path} that is obtained by assuming that the target moves with the mean velocity given by GP and then approximate $P_j\left(\mathbf{X}_{i}(k+1:k_{f})~|~\mathbf{X}_{i}(k)\right)$ along the nominal path.
In particular, define the sequence of nominal positions as $\hat{\mathbf{X}}_{ij}(k+1:k_{f})=[\hat{\mathbf{x}}_{ij}(k+1)\quad\dots\quad\hat{\mathbf{x}}_{ij}(k_{f})]$ where $\hat{\mathbf{x}}_{ij}(\tau+1)=\boldsymbol{\mu}_{ij}(\hat{\mathbf{x}}_{ij}(\tau))\Delta T+\hat{\mathbf{x}}_{ij}(\tau),\,\tau=k,\dots,k_{f}-1$, with the initial condition $\hat{\mathbf{x}}_{ij}(k)=\mathbf{x}_{i}(k)$. 
The velocity term $\boldsymbol{\mu}_{ij}(\hat{\mathbf{x}}_{ij}(\tau))$ is the mean vector computed using (\ref{eqn:gp_formula}). Then the pdf of the predicted trajectory is approximated as follows,
\begin{subequations}
	\label{eqn:approx_pos_pred_dist}
	\begin{align} & P_j\left(\mathbf{X}_{i}(k+1:k_{f})~|~\mathbf{X}_{i}(k)\right)\\
	& \approx\prod_{\tau=k}^{k_{f}-1}\int_{\mathbb{R}^{2}}\left[P\left(\mathbf{x}_{i}(\tau+1)~|~\mathbf{v}_{i}(\tau),\hat{\mathbf{x}}_{ij}(\tau)\right)\nonumber\right.\\
	&~~~~~~~~~\left.P_j\left(\mathbf{v}_{i}(\tau)~|~\hat{\mathbf{x}}_{ij}(\tau)\right)\right]d\mathbf{v}_{i}(\tau)\nonumber\\
	& =\prod_{\tau=k}^{k_{f}-1}P_j\left(\mathbf{v}_{i}(\tau)=\frac{\mathbf{x}_{i}(\tau+1)-\hat{\mathbf{x}}_{ij}(\tau)}{\Delta T}~|~\hat{\mathbf{x}}_{ij}(\tau)\right) \label{eqn:approx_pos_pred_dist1}
	\end{align}
\end{subequations}
where the equality (\ref{eqn:approx_pos_pred_dist1}) is obtained from the motion model (\ref{eqn:tar_kin_model}).
Equations (\ref{eqn:approx_pos_pred_dist})
indicates that the pdf of the predicted target trajectory can be approximated as the pdf of predicted velocities along the nominal path, and the resultant pdf is a product of Gaussian distributions. By simple algebraic manipulation, it can be shown that (\ref{eqn:approx_pos_pred_dist1}) is actually a Gaussian distribution $\mathcal{N}(\boldsymbol{\mu}_{ij,loc},\boldsymbol{\Sigma}_{ij,loc})$, where the mean vector $\boldsymbol{\mu}_{ij,loc}$ and the covariance matrix $\boldsymbol{\Sigma}_{ij,loc}$ is 
\begin{subequations}
	\label{eqn:loc_gp_mean_cov}
	\begin{align}
	\boldsymbol{\mu}_{ij,loc}&=\left[\hat{\mathbf{x}}_{ij}^T(k+1)\quad\dots\quad\hat{\mathbf{x}}_{ij}^T(k_f)\right]^T,\\ \boldsymbol{\Sigma}_{ij,loc}&=diag\left[\boldsymbol{\Sigma}_{ij}(\hat{\mathbf{x}}_{ij}(k+1))\quad\dots\quad\boldsymbol{\Sigma}_{ij}(\hat{\mathbf{x}}_{ij}(k_f))\right],
	\end{align}
\end{subequations}
where $diag$ means the block diagonal matrix. It is easy to see the mean is the vector of nominal positions.

\subsection{Decentralized Target Trajectory Fusion and Prediction}
Since the workspace is usually much larger than the sensing range of sensors, it can happen that each sensor only measures a subset of targets at each time step. 
To coordinate the sensing paths, it is important for sensors to fuse their local prediction to obtain a global consensus on targets' predicted trajectories. 
This subsection proposes a rumor-robust decentralized GP fusion approach.
Consider the fusion of $i$th target's prediction from sensors $j$ and $l$, where the pdfs of local prediction are $P_{j}\left(\mathbf{X}_{i}(k+1:k_{f})~|~\mathbf{X}_{i}(k)\right)$ and $P_{l}\left(\mathbf{X}_{i}(k+1:k_{f})~|~\mathbf{X}_{i}(k)\right)$, computed using (\ref{eqn:approx_pos_pred_dist}). 
The proposed fusion rule is as follows,
\begin{equation}
	\begin{aligned}\label{eqn:gp_fusion}
	&P\left(\mathbf{X}_{i}(k+1:k_{f})|\mathbf{X}_{i}(k)\right)\\
	&~~~~~~\propto P^{\beta_{j}w^{*}}\left(\mathbf{X}_{ij}(k+1:k_{f})|\mathbf{X}_{ij}(k)\right)\\
	&~~~~~~~~~~~P^{\beta_{l}(1-w^{*})}\left(\mathbf{X}_{il}(k+1:k_{f})|\mathbf{X}_{il}(k)\right),
	\end{aligned}
\end{equation}
where $\beta_{j}$ and $\beta_{l}$ are weighting factors that indicate each sensor's contribution to the combined prediction, and $w^{*}$ is the optimal weight based on Chernoff information \cite{nielsenChernoffInfoExpFamily11}. Following the strategy in \cite{caoGeneralizedPoEGPfusion14}, $\beta_{j}$ and $\beta_{l}$ are chosen as the difference in differential entropy between the prior and the posterior at $\mathbf{X}_{i}(k+1:k_{f})$ to ensure that the more information an agent contains about the $i$th target's prediction, the more it contributes to the combined prediction.
Using the fact that for a Gaussian distribution $P(\mathbf{x})\sim\mathcal{N}(\boldsymbol{\mu},\boldsymbol{\Sigma})$, its exponential is also a Gaussian distribution with a scaled covariance matrix \cite{caoGeneralizedPoEGPfusion14}, i.e. $P^{\alpha}(\mathbf{x})\sim\mathcal{N}(\boldsymbol{\mu},\alpha^{-1}\boldsymbol{\Sigma})$, the predictive mean and covariance matrix of $P\left(\mathbf{X}_{i}(k+1:k_{f})~|~\mathbf{X}_{i}(k)\right)$ are
\begin{equation}
\begin{aligned}
	\boldsymbol{\mu}_{i}^{*}&=\boldsymbol{\boldsymbol{\Sigma}}_{i}^{*}\left(\beta_{j}w^{*}\boldsymbol{\boldsymbol{\Sigma}}_{ij,loc}^{-1}\boldsymbol{\mu}_{ij,loc}+\beta_{l}(1-w^{*})\boldsymbol{\boldsymbol{\Sigma}}_{il,loc}^{-1}\boldsymbol{\mu}_{il,loc}\right)\\
	\boldsymbol{\boldsymbol{\Sigma}}_{i}^{*}&=\left(\beta_{j}w^{*}\boldsymbol{\boldsymbol{\Sigma}}_{ij,loc}^{-1}+\beta_{l}(1-w^{*})\boldsymbol{\boldsymbol{\Sigma}}_{il,loc}^{-1}\right)^{-1}
\end{aligned}
\label{eqn:gp_fusion_analy}
\end{equation}
where $\left(\boldsymbol{\mu}_{ij,loc},\boldsymbol{\boldsymbol{\Sigma}}_{ij,loc}\right)$ and $\left(\boldsymbol{\mu}_{il,loc},\boldsymbol{\boldsymbol{\Sigma}}_{il,loc}\right)$ are the mean and covariance pairs of sensor $j$ and $l$ 's local GP prediction.

The proposed fusion rule (\ref{eqn:gp_fusion}) is similar to the generalized product of GP experts (gPoE) method proposed in \cite{caoGeneralizedPoEGPfusion14}. However, PoE-based GP fusion approaches cannot be directly applied to sensor networks since these methods assume that the training data for each agent are disjoint.
However, in sensor networks, common information can exist in the training data of different sensors, such as a target is simultaneously measured by multiple sensors. 
Avoiding double-counting is therefore a key requirement for consistent data fusion in sensors networks \cite{campbellDistributedDataFusion16}. 
To avoid rumor propagation, this paper utilizes the weighted exponential product rule for data fusion. The Chernoff information is used to compute the optimal fusion weight, which has been shown effective to reduce rumor propagation in sensor networks \cite{ahmedFastChernoofFusionGM12}.
For two arbitrary pdfs $P_{a}(\mathbf{x})$ and $P_{b}(\mathbf{x})$,
the optimal Chernoff weight is obtained by minimizing their Chernoff
information, i.e.,
\todohere{figure out the definition of Chernoff information. seems quite confusing in different literature.}
\begin{equation*}
w^{*}=\arg\max_{w\in[0,1]}-\log\int\left[P_{a}(\mathbf{x})\right]^{w}\left[P_{b}(\mathbf{x})\right]^{1-w}d\mathbf{x}.
\end{equation*}
The combined pdf is then  $P(\mathbf{x})=\left[P_{a}(\mathbf{x})\right]^{w^{*}}\left[P_{b}(\mathbf{x})\right]^{1-w^{*}}$. 

The main difficulty of using Chernoff weight for information fusion is that for genenral distributions, there is usually no analytic expression for their Chernoff information, thus computing the optimal weight causes significant computational overhead \cite{ahmedFastChernoofFusionGM12}.
However, the Chernoff information for two multivariate Gaussian distributions, $P_{a}(\mathbf{x})\sim\mathcal{N}(\boldsymbol{\mu}_{a},\boldsymbol{\Sigma}_{a})$ and $P_{b}(\mathbf{x})\sim\mathcal{N}(\boldsymbol{\mu}_{b},\boldsymbol{\Sigma}_{b})$, can be expressed in the closed form \cite{nielsenChernoffInfoExpFamily11}, and the optimal Chernoff weight can be computed as follows,
\begin{align}
w^{*} & =\arg\min_{w\in[0,1]}\frac{1}{2}\log\frac{\left|w\boldsymbol{\Sigma}_{a}+(1-w)\boldsymbol{\Sigma}_{b}\right|}{|\boldsymbol{\Sigma}_{a}|^{w}|\boldsymbol{\Sigma}_{b}|^{1-w}}\label{eqn:opt_chern_weight}\\
 & +\frac{w(1-w)}{2}\left(\boldsymbol{\mu}_{a}-\boldsymbol{\mu}_{b}\right)^{T}\left(w\boldsymbol{\Sigma}_{a}+(1-w)\boldsymbol{\Sigma}_{b}\right)\left(\boldsymbol{\mu}_{a}-\boldsymbol{\mu}_{b}\right)\nonumber 
\end{align}
The optimal Chernoff weight, $w^{*}$, in (\ref{eqn:gp_fusion_analy}) can therefore be efficiently computed between the following two Gaussian distributions,
\begin{align*}
P^{\beta_{j}}\left(\mathbf{X}_{ij}(k+1:k_{f})~|~\mathbf{X}_{ij}(k)\right)&\sim\mathcal{N}(\boldsymbol{\mu}_{ij,loc},\beta_{j}^{-1}\boldsymbol{\Sigma}_{ij,loc}),\\
P^{\beta_{l}}\left(\mathbf{X}_{il}(k+1:k_{f})~|~\mathbf{X}_{il}(k)\right)&\sim\mathcal{N}(\boldsymbol{\mu}_{il,loc},\beta_{l}^{-1}\boldsymbol{\Sigma}_{il,loc}),
\end{align*}
using nonlinear optimization algorithms \cite{nocedalNumericalOpt06}.
Using (\ref{eqn:gp_fusion_analy}) and (\ref{eqn:opt_chern_weight}) for each pair of sensors along the tree-structured communication network, the rumor-robust decentralized GP fusion can be conducted efficiently. The fused prediction at the root sensor is propagated back to all sensors such that sensors have the same fused pdf, which we refer to as
$P_{fuse}\left(\mathbf{X}_{i}(k+1:k_{f})~|~\mathbf{X}_{i}(k)\right)$ for the $i$th target.

\section{Decentralized Sensor Planning}\label{sec:dec_ctrl}
This section presents the decentralized IPP algorithm used in RESIN, which uses the sequential planning strategy \cite{krauseNearOptObserveSelectSubmodFunc07,atanasovDecentralizedActiveInfo15,damesDetectingLocalizingTrackingFISST17}. As shown in \cite{krauseNearOptObserveSelectSubmodFunc07,atanasovDecentralizedActiveInfo15}, sequential planning is guaranteed to generate near-optimal solutions compared to the centralized planning when the objective function is submodular and monotonic, for which MI satisfies. In the sequential planning, given a planning order, each sensor first receives the planning information from its predecessors (Section \ref{subsec:fuse_pred_plan}), then it computes its own optimal path (Section \ref{subsec:local_obj}), and sends the new planning information to the next sensor in the sequence. 
The planning order can be pre-defined or determined online for each communication round.
Without loss of generality, we assume the planning order corresponds to sensors' indices in following analysis. The decentralized IPP algorithm is characterized by efficiency in both communication (Section \ref{subsec:fuse_pred_plan}) and computation (Section \ref{subsec:local_obj}).

\subsection{Fusing Predecessors' Plans}\label{subsec:fuse_pred_plan}
For the $j$th sensor, given the planned paths of the first $j-1$ sensors, $\mathbf{S}_{j-1}(k:k_{f})=[\mathbf{s}_{1}(k:k_{f})\quad\dots\quad\mathbf{s}_{j-1}(k:k_{f})]$,
the local planning problem becomes
\begin{equation}\label{eqn:local_ipp}
\begin{aligned}
\mathbf{u}_{j}^{*}\left(k:k_{f}\right) & =\arg\max_{\mathbf{u}_{j}\left(k:k_{f}\right)}J_{j}\left(\mathbf{u}_{j}\left(k:k_{f}\right);\mathbf{S}_{j-1}(k:k_{f})\right)\\
s.t.\; & \mathbf{s}_{j}(\tau+1)=\mathbf{g}\left(\mathbf{s}_{j}(\tau),\mathbf{u}_{j}(\tau)\right),\\
& \mathbf{s}_{j}(\tau)\in\mathcal{S},\,\mathbf{u}_{j}(\tau)\in\mathcal{U},\,\tau=k,\dots,k_{f}
\end{aligned}
\end{equation}
The objective function is defined as the mutual information between target prediction and $j$th sensor's planned path, conditioned on the first $j-1$ sensors' plans, i.e.,
\begin{multline*}
J_{j}\left(\mathbf{u}_{j}\left(k:k_{f}\right);\mathbf{S}_{j-1}(k:k_{f})\right)=\\
I\left(\mathbf{X}\left(k+1:k_{f}\right);\mathbf{z}_{j}\left(k+1:k_{f}\right)\vert\mathbf{Z}_{j-1}(k+1:k_{f})\right)
\end{multline*}%
where $\mathbf{Z}_{j-1}(k+1:k_{f})$ represents the predicted measurements of the first $j-1$ sensors' planned paths.


In order to fuse predecessor sensors' predicted measurements, the globally fused target prediction, $P_{fuse}\left(\mathbf{X}_{i}(k+1:k_{f})~|~\mathbf{X}_{i}(k)\right)$, is treated as the prior distribution of targets' prediction. Similar to Section \ref{subsec:local_gp_pred}, the Bayesian fusion approach is used to compute the posterior distribution conditioned on $\mathbf{Z}_{j-1}(k+1:k_{f})$. 
Define $\hat{\mathbf{X}}_{i}\left(k+1:k_{f}\right)=[\hat{\mathbf{x}}_i(k+1)\quad\dots\quad\hat{\mathbf{x}}_i(k_f)]$ as the nominal path obtained from $P_{fuse}\left(\mathbf{X}_{i}(k+1:k_{f})~|~\mathbf{X}_{i}(k)\right)$, the pdf after fusing $\mathbf{Z}_{j-1}(k+1:k_{f})$ is
\begin{subequations}
\label{eqn:gp_fusion_seq_prior}
\begin{align}
&P_{j,pre}\left(\mathbf{X}_{i}\left(k+1:k_{f}\right)~|~\mathbf{Z}_{j-1}(k+1:k_{f})\right)\nonumber\\
&\propto P_{fuse}\left(\mathbf{X}_{i}\left(k+1:k_{f}\right)\right)\nonumber\\
&~~~~~~\prod_{l=1}^{j-1}P\left(\mathbf{z}_{l}(k+1:k_{f})~|~\mathbf{X}_{i}\left(k+1:k_{f}\right)\right)\label{eqn:gp_fusion_seq_prior_factor}\\
&\approx\prod_{\tau=k}^{k_{f}-1}P\left(\mathbf{v}_{i}(\tau)=\frac{\mathbf{x}_{i}(\tau+1)-\hat{\mathbf{x}}_{i}(\tau)}{\Delta T}\right)\nonumber\\
&~~~~~~\prod_{l=1}^{j-1}P\left(\mathbf{z}_{l}(\tau)~|~\hat{\mathbf{x}}_{i}(\tau)\right)
\label{eqn:gp_fusion_seq_prior_approx}
\end{align}
\end{subequations}
The factorization in (\ref{eqn:gp_fusion_seq_prior_factor}) is obtained by the conditional independence of measurements from different sensors given the target positions.
Equation 
(\ref{eqn:gp_fusion_seq_prior_approx}) is obtained similar to that in (\ref{eqn:approx_pos_pred_dist}), where the pdf is approximated along the nominal path. 
The predicted measurement from $l$th sensor, $\mathbf{z}_{l}(\tau)$, is assumed to be nonempty if the nominal position $\hat{\mathbf{x}}_{i}(\tau)$ lies in the sensor's FOV at $\tau$. 

The prior $P\left(\mathbf{v}_{i}(\tau)=\frac{\mathbf{x}_{i}(\tau+1)-\hat{\mathbf{x}}_{i}(\tau)}{\Delta T}\right)$ can be directly obtained by marginalizing $P_{fuse}\left(\mathbf{X}_{i}(k+1:k_{f})~|~\mathbf{X}_{i}(k)\right)$ over all time steps except $\tau$, and can be easily shown to be a Gaussian distribution, denoted as $\mathcal{N}(\boldsymbol{\mu}_{i,fuse}(\tau),\boldsymbol{\Sigma}_{i,fuse}(\tau))$. Since the measurement model is linear Gaussian, an analytical expression of $P_{j,pre}\left(\mathbf{X}_{i}\left(k+1:k_{f}\right)~|~\mathbf{Z}_{j-1}(k+1:k_{f})\right)$ can be obtained. 
In particular, let $\mathbb{I}\{\hat{\mathbf{x}}_{i}(\tau)\in\mathcal{F}\left(\mathbf{s}_{l}(\tau)\right)\}$ represent the indicator function and it equals $1$ if and only if the the nomial position
$\hat{\mathbf{x}}_{i}(\tau)$ lies in the sensor's planned FOV at $\tau$.
Then it can be derived, using the conjugacy property of Gaussian prior and likelihood functions \cite{bishopPRML06}, that given the prior covariance matrix $\boldsymbol{\Sigma}_{i,fuse}(\tau)$ and let $n(\tau)=\sum_{l=1}^{j-1}\mathbb{I}\{\hat{\mathbf{x}}_{i}(\tau)\in\mathcal{F}\left(\mathbf{s}_{l}(\tau)\right)\}$ represent the number of sensors in the first $j-1$ sensors that can measure the $i$th target at time $\tau$, then the posterior covariance is
\begin{equation*}
\boldsymbol{\Sigma}_{ij,pre}(\tau)=\left(\boldsymbol{\Sigma}_{i,fuse}^{-1}(\tau)+n(\tau)\boldsymbol{\Sigma}_{\varepsilon}^{-1}\right)^{-1},\,\tau=k+1,\dots,k_f.
\end{equation*} 
Therefore, the fused pdf can be compactly represented as
\begin{multline*}
    P_{j,pre}\left(\mathbf{X}_{i}\left(k+1:k_{f}\right)~|~\mathbf{Z}_{j-1}(k+1:k_{f})\right)\\
    \sim\mathcal{N}\left(\boldsymbol{\mu}_{ij,pre}(k+1:k_{f}),\boldsymbol{\Sigma}_{ij,pre}(k+1:k_{f})\right),
\end{multline*}
where the covariance matrix $\boldsymbol{\Sigma}_{ij,pre}(k+1:k_{f})$ is
{\small\begin{equation*}
\boldsymbol{\Sigma}_{ij,pre}\left(k+1:k_{f}\right)=diag\left[\boldsymbol{\Sigma}_{ij,pre}(k+1)\quad\dots\quad\boldsymbol{\Sigma}_{ij,pre}(k_f)\right].
\end{equation*}}%
Since the objective function only depends on covariance matrix, we ignore the expression of mean $\boldsymbol{\mu}_{ij,pre}(k+1:k_{f})$. 

The key observation here is that, to compute $\boldsymbol{\Sigma}_{ij,pre}\left(k+1:k_{f}\right)$, the only information needed from predecessor sensors is the times that each sensor can expect to detect the target in the planning interval, i.e., $n(t)$. Since all sensors share the same nominal path of the target, $\hat{\mathbf{X}}_{i}\left(k:k_{f}\right)$, thanks to the decentralized GP fusion, each sensor only needs to receive the counting from its predecessor, then add its own measurement times to the total counting, and send the updated counting to the next sensor. The communication overhead between each pair of sensors is therefore constant and independent of the  number of predecessor sensors. In contrast, in state-of-the-art sequential planning approaches \cite{krauseNearOptObserveSelectSubmodFunc07,atanasovDecentralizedActiveInfo15,damesDetectingLocalizingTrackingFISST17}, the transmitted information to each sensor is the planned path from all predecessor sensors, which has the communication burden of $O(N)$. This shows our sequential planning approach significantly reduces the communication overhead compared to the state-of-the-art.


\subsection{Local Objective Function}\label{subsec:local_obj}
The fused pdf $P_{j,pre}\left(\mathbf{X}_{i}\left(k:k_{f}\right)~|~\mathbf{Z}_{j-1}(k:k_{f})\right)$ is now used as the prior pdf for $j$th sensor's path planning. Given the $j$th sensor's future control inputs $\mathbf{u}_{j}\left(k:k_{f}\right)$ and the consequent future measurements $\mathbf{z}_{j}\left(k:k_{f}\right)$, the posterior pdf can be obtained similar to (\ref{eqn:gp_fusion_seq_prior}), i.e.,
\begin{align}
    ~&P_{j,plan}\left(\mathbf{X}_{i}\left(k+1:k_{f}\right)~|~\mathbf{z}_{j}\left(k+1:k_{f}\right),\mathbf{Z}_{j-1}(k+1:k_{f})\right)\nonumber\\
~&\propto P_{j,pre}\left(\mathbf{X}_{i}\left(k+1:k_{f}\right)|\mathbf{Z}_{j-1}(k+1:k_{f})\right)\label{eqn:plan_dist}\\
&\qquad P\left(\mathbf{z}_{j}\left(k+1:k_{f}\right)|\mathbf{X}_{i}\left(k+1:k_{f}\right)\right)\nonumber\\
~&\approx\prod_{\tau=k+1}^{k_{f}}P\left(\mathbf{z}_{j}(\tau)~|~\hat{\mathbf{x}}_{i}(\tau)\right)\nonumber\\
&~~~~~~P\left(\hat{\mathbf{v}}_{i}(\tau)=\frac{\mathbf{x}_{i}(\tau+1)-\hat{\mathbf{x}}_{i}(\tau)}{\Delta T}~|~\mathbf{Z}_{j-1}(k+1:k_{f})\right) \nonumber\\
&\sim\mathcal{N}\left(\boldsymbol{\mu}_{ij,plan}(k+1:k_{f}),\boldsymbol{\Sigma}_{ij,plan}(k+1:k_{f})\right),\nonumber
\end{align}
where the covariance matrix is 
\begin{multline*}
	\boldsymbol{\Sigma}_{ij,plan}\left(k+1:k_{f}\right)=diag\left[\boldsymbol{\Sigma}_{ij,plan}(k+1)\quad\dots\quad\right.\\
	\left.\boldsymbol{\Sigma}_{ij,plan}(k_f)\right].
\end{multline*}
Again, using the conjugacy of Gaussian distribution, the covariance matrix of the posterior pdf can be computed in a closed-form, i.e., for $\tau=k+1,\dots,k_{f}$,
\begin{equation}
\boldsymbol{\Sigma}_{ij,plan}(\tau)=\left(\boldsymbol{\Sigma}_{ij,pre}^{-1}(\tau)+\mathbb{I}\{\hat{\mathbf{x}}_{i}(\tau)\in\mathcal{F}\left(\mathbf{s}_{j}(\tau)\right)\}\boldsymbol{\Sigma}_{\varepsilon}^{-1}\right)^{-1}.\label{eqn:gp_fusion_seq_post_cov}
\end{equation}


Now we derive the closed-form of the objective function. Notice that 
{\small\begin{subequations}
\label{eqn:dec_ctrl_obj_simp}
\begin{align}
    ~&J_{j}\left(\mathbf{u}_{j}\left(k+1:k_{f}\right)~|~\mathbf{S}_{j-1}(k+1:k_{f})\right)\nonumber\\
    &=\sum_{i=1}^{M}I\left(\mathbf{X}_{i}\left(k+1:k_{f}\right);\mathbf{z}_{j}\left(k+1:k_{f}\right)~|~\mathbf{Z}_{j-1}(k+1:k_{f})\right)\label{eqn:dec_ctrl_obj_simp_decomp1}\\
&=\sum_{i=1}^{M}\sum_{\tau=k+1}^{k_{f}}I\left(\mathbf{X}_{i}\left(\tau\right);\mathbf{z}_{j}\left(\tau\right)~|~\mathbf{Z}_{j-1}(k+1:k_{f})\right)\label{eqn:dec_ctrl_obj_simp_decomp2}
\end{align}
\end{subequations}}%
where (\ref{eqn:dec_ctrl_obj_simp_decomp1}) is due to the independence of GP models for different targets and (\ref{eqn:dec_ctrl_obj_simp_decomp2}) is due to the block diagonal shape of $\boldsymbol{\Sigma}_{ij,plan}$.
Given the analytic expression of MI between Gaussian distributions \cite{bishopPRML06}, it can be derived that
\begin{multline*}
J_{j}\left(\mathbf{u}_{j}^{*}\left(k:k_{f}\right)~|~\mathbf{S}_{j-1}(k:k_{f})\right)=\\\sum_{i=1}^{M}\sum_{\tau=k}^{k_{f}}\frac{1}{2}\log\det\frac{\boldsymbol{\Sigma}_{ij,pre}(\tau)}{\boldsymbol{\Sigma}_{ij,plan}(\tau)}.
\end{multline*}

The indicator function in (\ref{eqn:gp_fusion_seq_post_cov}) makes the IPP problem a mixed integer nonlinear programming problem, which is notoriously difficult to solve. 
To overcome this difficulty, we consider an approximate objective function, where the indicator function is replaced by the constant $1$ and a weighting factor is added to the MI at each step.
In particular, define the weighting factor
\begin{multline*}
    \psi(\tau)=\max\left(0,\right.\\
    \left.1-\frac{\left(\|[s_{j,x}(\tau),s_{j,y}(\tau)]^{T}-\hat{\mathbf{x}}_{i}\left(\tau\right)\|_{2}-\frac{r_{j}}{2}\right)^{2}}{(\frac{r_{j}}{2})^{2}}\right).
\end{multline*}
Then the objective function is defined as
{\small\begin{multline}
J_{j}\left(\mathbf{u}_{j}\left(k:k_{f}\right)~|~\mathbf{S}_{j-1}(k:k_{f})\right)\\
=\sum_{i=1}^{M}\sum_{\tau=k}^{k_{f}}\frac{\psi(\tau)}{2}\log\det\frac{\boldsymbol{\Sigma}_{ij,pre}(\tau)}{\tilde{\boldsymbol{\Sigma}}_{ij,plan}(\tau)}
\end{multline}}%
where $\tilde{\boldsymbol{\Sigma}}_{ij,plan}(\tau)=\left(\boldsymbol{\Sigma}_{ij,pre}^{-1}(\tau)+\boldsymbol{\Sigma}_{\varepsilon}^{-1}\right)^{-1}$.
Using this closed-form objective function, the decentralized IPP problem (\ref{eqn:local_ipp}) can be efficiently solved using nonlinear optimization algorithms.
\todohere{prove submodularity}

\section{Simulation Setup and Results}
\label{sec:sim}
Two simulations are conducted to evaluate the effectiveness of RESIN.
The first simulation evaluates the decentralized GP learning and fusion in RESIN by comparing it to the centralized GP and the GP without fusion methods. Stationary sensors are used to avoid the influence of planning algorithms.
The second simulation subsequently considers mobile sensors and compares RESIN to the centralized GP planner, random planner, and the nearest target following planner. 




\subsection{Evaluating Decentralized GP Learning and Fusion}
In this simulation, four stationary sensors are randomly placed in a $10m \times 10m$ workspace.
A total of eight targets enter the workspace, each of which has a different trajectory. The sensor's sensing range is $r_j=5m,\,\forall j=1,\dots,N$. The decentralized GP learning and fusion in RESIN is compared to the centralized GP and the GP without fusion methods. In the centralized GP method, all sensors share their measurements with other sensors such that the GP learning and prediction are based on all sensor measurements. For the GP without fusion method, sensors do not communicate with each other and therefore the local GP learning and prediction is based on the sensor's local measurements. 

\begin{figure}
\includegraphics[width=0.5\textwidth]{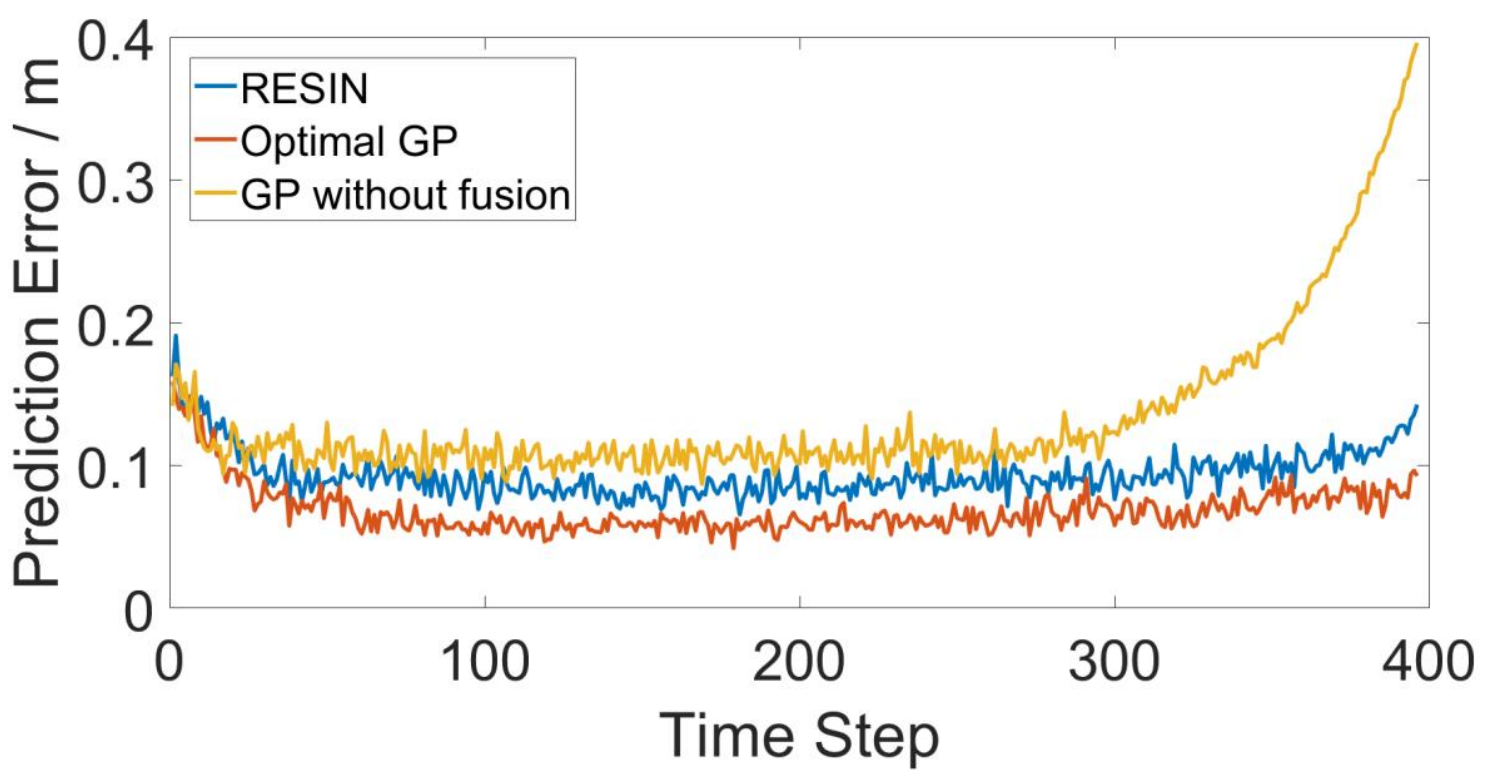}
\centering
\caption{Average prediction error of targets using different GP fusion strategies.}
\label{fig:error_fixed}
\end{figure}

Figure \ref{fig:error_fixed} compares the average prediction error of all sensors' predicted trajectory of every target in the planning interval under different prediction approaches. The planning horizon is five steps. As expected, the centralized GP has the minimum prediction error among all three approaches, as the sensors have direct access to all measurements and therefore is able to make the most accurate prediction. RESIN has a slightly larger prediction error than the centralized GP fusion. In contrast, GP without fusion leads to the worst prediction error since each sensor makes prediction only based on its own measurements. 
The simulation result shows that RESIN is an effective fusion approach that achieves similar performance as that of the centralized GP, while significantly reducing the computational complexity of learning and prediction compared to the centralized GP.

\subsection{Evaluating RESIN}
In this simulation, four mobile sensors are randomly placed in a $30m\times 30m$ workspace. 
There is a total of eight moving targets and each target moves in a different pattern. 
The velocity of each sensor is bounded in the range of $[0, 3](m/s)$ and the control input of each robot is also bounded, defined as follows,
\begin{equation*}
   \begin{bmatrix} -\frac{\pi}{6} \\ -5 (m / s) \end{bmatrix} \leq \mathbf{u} \leq \begin{bmatrix} \frac{\pi}{6} \\ 5 (m / s) \end{bmatrix}.
\end{equation*}
The sensing range of the sensor's camera is $r_j=5m,\,\forall j=1,\dots,N$. The planning horizon is five steps. RESIN is compared to three benchmark methods, including the centralized GP planner, nearest target following planner, and the random planner. In the centralized GP planner, the GP learning, prediction, and planning are conducted in a centralized way, where all sensors' measurements are shared, and the planning is conducted for all sensors at the same time. 
The nearest target following planner drives each sensor to pursue the closest target based on its locally estimated target position. The random planner generates random control inputs for each sensor.

\begin{figure}
    \includegraphics[width=0.48\textwidth]{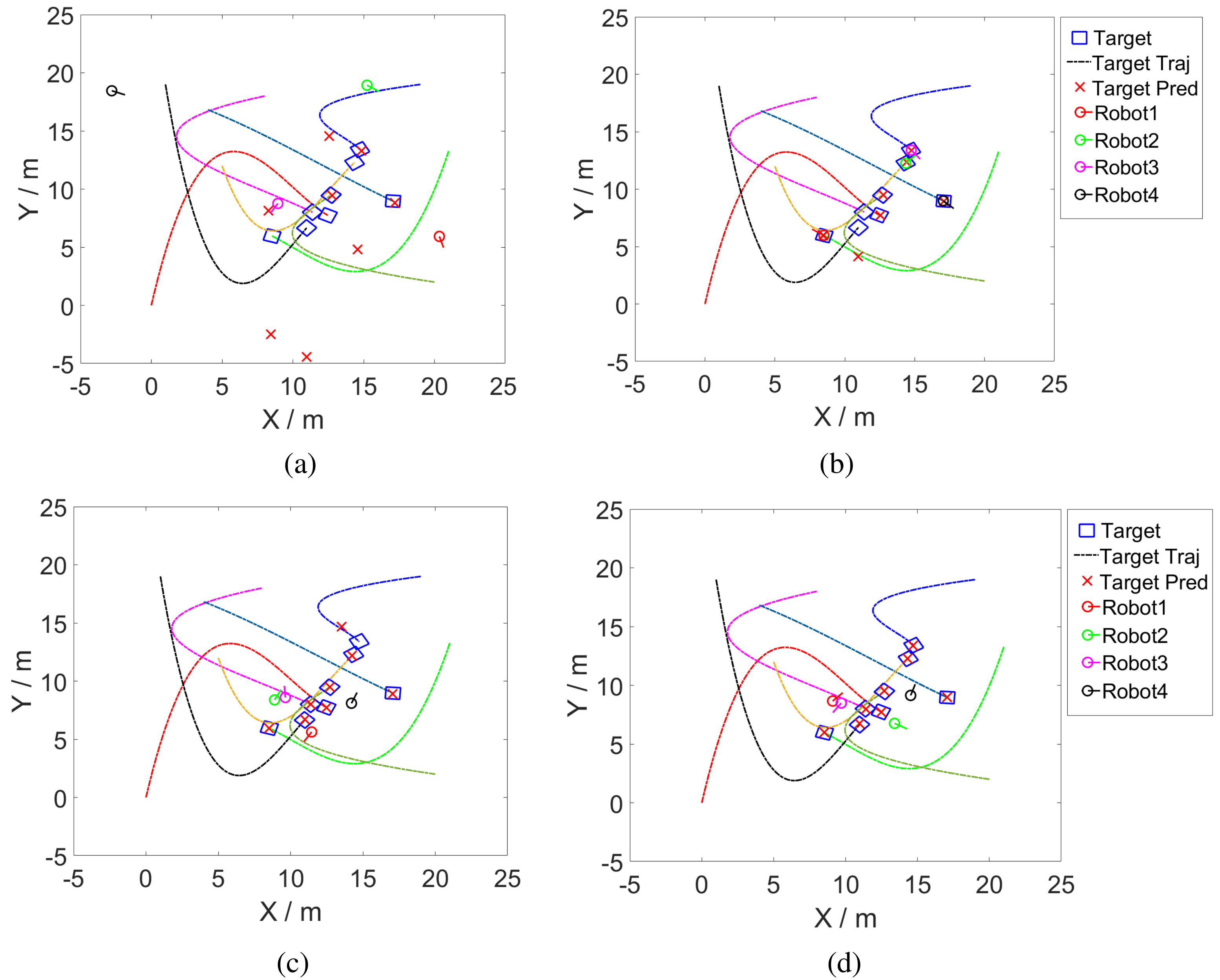}
    \centering
    \caption{Target prediction and tracking performance using (a) random planner, (b) nearest target following planner, (c) RESIN, and (d) centralized GP planner.}
\label{fig:tracking}
\end{figure}

\begin{figure}
	\includegraphics[width=0.47\textwidth]{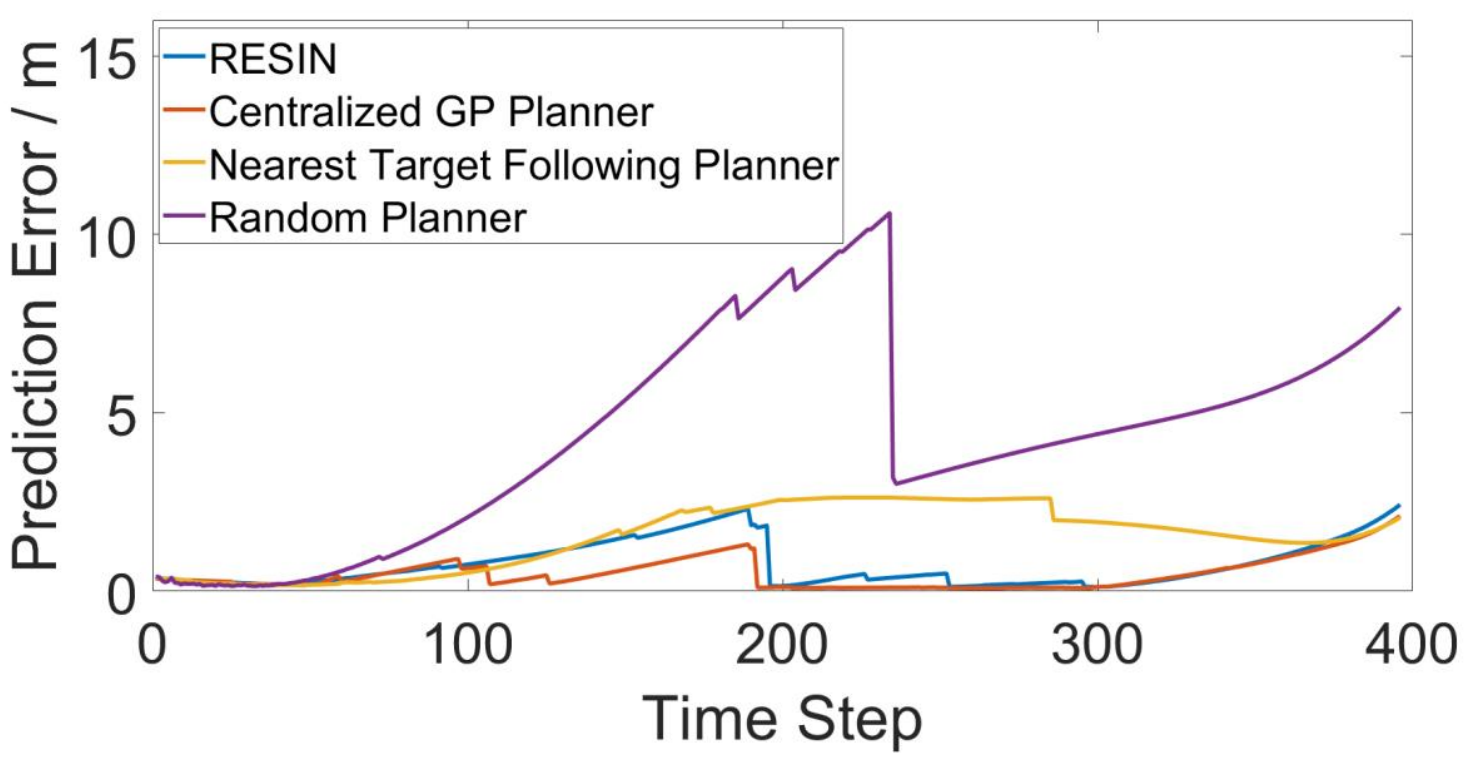}
	\centering
	\caption{Average prediction error of targets using different fusion and planning strategies.}
	\label{fig:error}
\end{figure}

Figure \ref{fig:tracking} visually compares the performance of these four planners. As expected, the centralized GP planner achieves the best performance as it has access to all measurements and can simultaneously plan for all sensors. The performance of RESIN is very similar to that of the centralized GP planner, as sensors can closely keep track of targets and make accurate prediction of target positions. In contrast, under the nearest target following planner, sensors lose track of several targets, as the target prediction becomes inaccurate. The problem is even exacerbated with the random planner, as the sensors do not take into account the target positions and lose track of targets very soon. 

This observation is further supported in Figure \ref{fig:error}, in which the average prediction error of all sensors' predicted trajectory of every target under different planning strategies are quantitatively compared. RESIN outperforms both the nearest target following planner and the random planner in general, and has very similar performance as that of the centralized GP planner.

\section{Conclusion}\label{sec:conclusion}
This paper proposes RESIN for sensor networks to actively learn GP motion models of moving targets. Characterized by the computational and communication efficiency, and the robustness to rumor propagation, RESIN is a powerful framework for mobile sensor networks. The future work will investigate the data association issue in the decentralized fusion. Combining the decentralized classification of target motion patterns with decentralized control is also a promising topic to study. Last, physical experiments using actual mobile sensors and targets will be conducted to evaluate its practical use.

\bibliographystyle{IEEEtran}
\bibliography{decentralizedCtrl_ref}
\end{document}